\documentclass[preprint,12pt]{elsarticle}
\usepackage{amssymb}
\journal{Astroparticle Physics}
\newcommand{\be}{\begin{equation}}
\newcommand{\ee}{\end{equation}}
\newcommand{\simless}{\lower.5ex\hbox{$\; \buildrel < \over \sim\;$}}
\newcommand{\simgreat}{\lower.5ex\hbox{$\; \buildrel > \over \sim\;$}} 
\newcommand{\mpro}{ m_{\rm p}}

\newcommand{\mesa}{\small{\sl MESA}} 
\newcommand{\delres}{\Delta{E}_R}
\newcommand{\metal}{{\cal Z}} 
\newcommand{\tend}{ t_{\rm end} } 
\font\Fermifont=cmssq8 scaled 1200

\begin{document}

\begin{frontmatter} 

\title{{$\qquad$ $\qquad$ $\qquad$ $\qquad$ $\qquad$ $\qquad$ 
$\qquad$ $\qquad$ {\Fermifont  LA-UR-18-27626} \\} 
{\bf Sensitivity of Carbon and Oxygen Yields to \\
the Triple-Alpha Resonance in Massive Stars}}  

\author[a]{Lillian Huang} 
\author[a,b]{Fred C. Adams}
\author[c,d]{and Evan Grohs}

\address[a]{Physics Department, University of Michigan, Ann Arbor, MI 48109} 
\address[b]{Astronomy Department, University of Michigan, Ann Arbor, MI 48109} 
\address[c]{Department of Physics, University of California Berkeley, 
Berkeley, California 94720} 
\address[d]{Theoretical Division, Los Alamos National Laboratory, 
Los Alamos, New Mexico 87545}


\begin{abstract}
Motivated by the possible existence of other universes, this paper
considers the evolution of massive stars with different values for the
fundamental constants. We focus on variations in the triple alpha
resonance energy and study its effects on the resulting abundances of
$^{12}$C, $^{16}$O, and larger nuclei. In our universe, the $0^{+}$
energy level of carbon supports a resonant nuclear reaction that
dominates carbon synthesis in stellar cores and accounts for the
observed cosmic abundances. Here we define $\delres$ to be the change
in this resonant energy level, and show how different values affect
the cosmic abundances of the intermediate alpha elements.  Using the
state of the art computational package {\small{\sl MESA}}, we carry
out stellar evolution calculations for massive stars in the range
$M_\ast$ = $15-40M_\odot$, and for a wide range of resonance
energies. We also include both solar and low metallicity initial
conditions. For negative $\delres$, carbon yields are increased
relative to standard stellar models, and such universes remain viable
as long as the production of carbon nuclei remains energetically
favorable, and stars remain stable, down to $\delres\approx-300$
keV. For positive $\delres$, carbon yields decrease, but significant
abundances can be produced for resonance energy increments up to
$\delres\approx+500$ keV. Oxygen yields tend to be anti-correlated
with those of carbon, and the allowed range in $\delres$ is somewhat
smaller. We also present yields for neon, magnesium, and silicon. With
updated stellar evolution models and a more comprehensive survey of
parameter space, these results indicate that the range of viable
universes is larger than suggested by earlier studies.
\end{abstract}  


\end{frontmatter}


\section{Introduction} 
\label{sec:intro}  

Over the past few decades, a detailed paradigm for the evolution of
our universe has been developed, and this framework provides a
successful explanation for many observed cosmic features (e.g., see
\cite{particlegroup} for a recent review). Some versions of this
theory also argue that our universe could be one portion of a much
larger region of space-time, sometimes known as the ``multiverse,'' 
i.e., our local region could represent one member of a vast collection
of other universes
\cite{carrellis,davies2004,deutsch,donoghue,ellis2004,lindemultiverse}.
Moreover, these alternate universes could have different realizations
of the laws of physics. Specifically, the constants of nature,
including the strengths of the fundamental forces and the masses of
the fundamental particles, could vary from region to region
\cite{dirac,tegmark,reessix,schellekens}. Many authors have studied
the effects of these possible variations in the laws of physics and
find that only certain ranges for the parameters allow for universes
to form cosmic structure and support working stars
\cite{carr,carter,bartip,hogan,aguirre,barnes2012,liviorees2018}. 
Related work studies the possible time variations of the laws of
physics in our universe \cite{barrow,uzan}. Since carbon is generally
considered a prerequisite for biology --- at least for life in
familiar forms --- the synthesis of carbon provides an important
constraint for habitable universes.

Carbon production in stellar interiors takes place through a rather 
complicated nuclear process known as the triple alpha reaction
\cite{clayton,kippenhahn,hansen}. Because of the intricate landscape
of nuclear binding energies and reaction rates, this process takes
place through a resonant reaction, which is enabled by a particular
excited state of the carbon nucleus. The resulting reaction rate for
carbon production depends sensitively on the value of the resonant
energy level (for greater detail, see Section \ref{sec:nuke} and
references therein). As a result, if the laws of physics take
different forms in other regions of space-time, the resonance energy
level could be different, and the amount of carbon produced would vary
accordingly. The objective of this paper is to determine how
variations in the triple alpha resonance energy affect the abundances
of the alpha elements produced in massive stars, with a focus on
$^{12}$C and $^{16}$O. The overall goal is to specify the range in
resonance energy, characterized the change $\delres$ (see equation
[\ref{delresdef}]), that allows the universe to be viable.

Possible variations to the energy level of the carbon resonance, and
their effects on stellar evolution, have been explored previously
\cite{livio,oberhummer,schlattl,ekstrom}. This paper generalizes
earlier work by taking advantage of continuing developments in
computational capabilities. This contribution uses the stellar
evolution package \mesa~(Modules for Experiments in Stellar
Astrophysics), a state of the art coding package that has been
recently developed for a host of applications and is publicly
available \cite{mesaone,mesatwo}. The standard \mesa~package does not
include changes to the triple alpha resonance level, so these
modifications must be implemented at the code-level. In addition to
using an updated stellar evolution code, this work also explores a
much larger regime of parameter space.  Whereas previous papers were
limited to relatively few values for the stellar mass $M_\ast$ and the
resonance energy increment $\delres$, this work considers much wider
ranges for $M_\ast$ and $\delres$, along different choices for the
stellar metallicity $\metal$ (our results are broadly consistent with
earlier work \cite{livio,oberhummer,schlattl} for given parameter
values). In addition to considering carbon and oxygen, we also compile
yields for larger alpha elements (neon, magnesium, and silicon).
Altogether, this paper reports the results from $\sim2400$ stellar
evolution simulations.

In assessing changes to the triple alpha process, the standard
approach, which we also follow, is to vary the energy level of the
$^{12}$C nucleus, but keep all other parameters the same. At the
fundamental level, however, variations in the excited state of nuclei
are determined by changes in the strengths of the fundamental forces,
especially the strong and electromagnetic interactions
\cite{epelbaum,epelbaum2011,epelbaum2012,lahde,meissner}.  
In principle, changes in these interaction strengths would affect all
nuclear characteristics, including binding energies and reaction
rates, not just the energy level of the $^{12}$C resonance of interest
here. In practice, however, small changes to the resonance energy lead
to large changes in carbon production.  Because of this extreme
sensitivity, we can implement variations to the carbon resonance
energy while keeping other nuclear parameters fixed. More
specifically, the required changes to the resonance energy are of
order 300 keV, whereas the binding energies of the nuclei are much
larger and fall in the range 28 -- 92 MeV.

This paper is organized as follows. We start with a brief review of
the triple alpha reaction in Section \ref{sec:nuke}, which also shows
how changes to the process are implemented. The stellar evolution
calculations are presented in Section \ref{sec:evolution}, including a
description of numerical considerations, basic evolution for massive
stars, the effects of changing the triple alpha resonance energy, and
the resulting carbon and oxygen yields over a wide range of parameter
space. The abundance of carbon required for a viable universe is
addressed in Section \ref{sec:context}, along with the possibility
that $^8$Be can be stable and obviate the need for the triple alpha
process. The relationship between the triple alpha resonance energy
and the fundamental parameters of particle physics is briefly
discussed. The paper concludes, in Section \ref{sec:conclude}, with 
a summary of our results and a discussion of their implications. 

\section{The Triple Alpha Reaction} 
\label{sec:nuke}  

After a star burns through the hydrogen fuel in its central core,
which is then composed primarily of helium, the star adjusts its
internal structure. The central regions condense so that core becomes
hotter and denser. Under these conditions, helium becomes the stellar
fuel and leads to the production of heavier elements. Given the tight
binding energy of helium nuclei --- alpha particles --- the natural
progression is for the helium nuclei to fuse together to synthesize
the so-called alpha elements: carbon, oxygen, and neon. In fact, after
hydrogen and helium, these three species are the most abundant
elements in our universe \cite{cameron,trimble}, with magnesium, 
silicon, and sulfur close behind. 

This nuclear chain is complicated by the fact that $^8$Be (and all
other nuclei with atomic mass number $A=8$) are unstable in our
universe. In the absence of stable $^8$Be, which provides a stepping
stone on the path to heavier alpha elements, the fusion of helium
takes place through the triple alpha reaction
\cite{clayton,kippenhahn,hansen}, where three helium nuclei combine to
make carbon. The net result of this process can be written in the form
\be
3\,\left(^4{\rm He}\right) \to \, {}^{12}\rm{C} + \gamma \,, 
\label{alpha3net} 
\ee
but intermediate steps are required.  In order to facilitate the
reaction (\ref{alpha3net}), the stellar core maintains a transient
population of unstable $^8$Be nuclei \cite{salpeter}. In this 
setting, the alpha particles fuse to produce $^8$Be, which decays back
into its constituent alpha particles with a half-life of approximately
$\tau_{1/2}\sim10^{-16}$ sec, 
\be
^4{\rm He} + \, ^4{\rm He} \longleftrightarrow  \, ^8{\rm Be}, 
\label{hehe2be} 
\ee
The forward reactions occur fast enough that the stellar core
maintains nuclear statistical equilibrium (NSE), which determines the
abundances of the relevant nuclear species. The resulting transient
population of $^8$Be is large enough for some of these unstable nuclei
to interact during their short lifetimes through the reaction 
\be
^4{\rm He} \, + {}^8{\rm Be} \to \, {}^{12}\rm{C} \,.
\label{hebe2c} 
\ee
Given the densities and temperatures of helium-dominated stellar
cores, the non-resonant reaction does not take place fast enough to
explain the observed carbon and oxygen abundances found in our
universe. However, the $^{12}$C nucleus has an excited state at an
accessible energy so that this reaction can operate in a resonant
manner, which increases the reaction rate and allows stars to produce
the observed cosmic abundances of carbon. Both the existence and the
particular energy level of this excited state were predicted by Hoyle
\cite{hoyle}, and subsequent laboratory experiments \cite{dunbar}
measured the resonance with the anticipated properties (see also the
reviews of \cite{fowler,wallerstein}). The relevant excited state 
has an energy of 7.6444 MeV, and corresponds to a $0^+$ nuclear 
state of the $^{12}$C nucleus. Significantly, this energy is somewhat
larger than the energy of the alpha particle and the $^8$Be nucleus
considered as separate particles (see equation [\ref{hebe2c}]). The
efficacy of carbon production is highly sensitive to the energy of
this resonance.

As outlined above, the production of carbon relies, in part, on the
intermediate reaction from equation (\ref{hehe2be}), even though the
product $^8$Be is unstable. The reaction rate for this process
\cite{clayton,kippenhahn,epelbaum} depends on the energy difference 
\be
(\Delta E)_b \equiv E_8 - 2 E_4 \,. 
\ee
The ground state energies of $^4$He and $^8$Be are denoted as $E_4$
and $E_8$, respectively. Similarly, the ground state of carbon is
denoted here as $E_{12}$ and the excited state (the $0^+$ resonance) 
is $E_{12}^\star$. In the reaction (\ref{hebe2c}), the energy 
difference between the excited carbon nucleus and the interacting 
nuclei is then given by  
\be
(\Delta E)_h = E_{12}^\star - E_8 - E_4 \,. 
\ee
The energy scale $E_R$ of the resonant reaction can then be 
defined as follows: 
\be
E_R \equiv (\Delta E)_b + (\Delta E)_h = E_{12}^\star - 3E_4 \,. 
\label{resenergy} 
\ee
This energy level has been experimentally measured to be
$(E_R)_0\approx379.5$ keV. Given the above definitions, the 
resonant reaction rate $R_{3\alpha}$ for the triple alpha 
process at temperature $T$ can be written in the form 
\be
R_{3\alpha} = 3^{3/2} n_\alpha^3 
\left( {2\pi\hbar^2 \over |E_4|kT} \right)^3 
{\Gamma_\gamma \over \hbar} 
\exp \left[ - {E_R\over kT} \right] \,,
\label{trialpharate} 
\ee
where $n_\alpha$ is the number density of alpha particles and
$\Gamma_\gamma\approx0.0037$ eV is the radiative width of the Hoyle 
state \cite{nomoto}. Note that the energy scale $E_R$ appears in 
the exponential term. As a result, the net reaction rate for carbon 
production is exponentially sensitive to the value of $E_R$. Notice
also that the argument of the exponential function varies as $T^{-1}$.
For non-resonant reactions, Coulomb barrier penetration convolved with
a thermal distribution of particle velocities leads to the usual
$T^{-1/3}$ dependence of the reaction rate
\cite{clayton,kippenhahn,hansen}; for resonant reactions, the
convolution picks out a single energy (here $E_R=E_{12}^\star-3E_4$)
and hence the factor $\exp[-E_R/kT]$.

In this treatment, we allow the energy of the resonance to take
different values than that of our universe. Specifically, we define
the energy increment 
\be
\delres \equiv E_R - (E_R)_0 \,,
\label{delresdef} 
\ee
where the subscript `0' denotes the value in our universe. The 
energy difference $\delres$ represents the most important 
variable in the problem. 

Equation (\ref{trialpharate}) shows that higher energies for the
$^{12}$C resonance lead to suppression of the triple alpha
reaction. Specifically, for a given temperature, increasing the
resonance level ($\delres>0$) results in a lower reaction rate for
carbon production (helium burning). However, the star must generate
enough energy to produce the pressure necessary to support itself
against gravity. For $\delres>0$, the stellar core must increase its 
temperature to compensate. Since the reaction rate is exponentially
sensitive to the central temperature, only modest increases are 
necessary. However, these higher temperatures allow carbon nuclei 
in the core to fuse into oxygen through the reaction 
\be
^4{\rm He} \, + {}^{12}\rm{C} \to \, {}^{16}\rm{O} \,.
\label{hec2o} 
\ee
In this reaction, the $^{16}$O nucleus has an energy level of 7.1187
MeV, which is below the combined energy of the reactants $^{12}$C and
$^{4}$He (which have energy 7.1616 MeV). The reaction (\ref{hec2o})
thus takes place in a non-resonant manner. In addition, a sizable
Coulomb barrier between the reacting nuclei must be overcome, and this
barrier depends sensitively on temperature. These features allow some 
fraction of the carbon to survive. 

The basic problem that the triple alpha reaction poses for carbon
production can be stated as follows: If the resonance energy level is
raised $(\delres>0)$, then the temperature required for carbon
production increases. But this hotter temperature also increases the
rate of carbon depletion via equation (\ref{hec2o}). As a result, for
sufficiently large increases in $\delres$ --- and hence higher
operating temperatures --- the carbon produced by stellar
nucleosynthesis can be immediately be transformed into oxygen and
heavier elements. In this regime, relatively little carbon would be
left behind for making life forms and other interesting structures.

The extreme sensitivity of the triple alpha reaction rate to both
temperature and the resonance energy represents an important aspect of
the problem.  This sensitivity is often characterized by the indices 
defined by
\be
\Xi_T \equiv {T \over R_{3\alpha}} {d R_{3\alpha} \over dT} = 
- 3 + {E_R \over kT} \,,
\ee
where the second equality uses equation (\ref{trialpharate}) to 
evaluate the index, and 
\be
\Xi_E \equiv {\delres \over R_{3\alpha}} 
{d R_{3\alpha} \over d\delres} = - {\delres \over kT} \,. 
\ee
For our universe, $E_R\approx380$ keV, whereas typical operating
conditions for carbon production correspond to temperatures
$kT\approx9-17$ keV. As a result, the index $\Xi_T\approx20-40$, so
that the triple alpha reaction rate varies rapidly with changes in
temperature. Stellar cores thus have temperatures of order 
$kT\sim10$ keV during helium burning. Stability of the star requires
that the index $\Xi_T>0$, so we obtain a lower bound on the possible
values of the resonance energy, equivalently, on the scale $E_R$:
\be
E_R > 3 (kT) \simgreat 30\,{\rm keV} \,.  
\label{resconstraint} 
\ee
In other words, the resonance energy cannot be made arbitrarily small,
but this correction ($\sim30$ keV) is small compared to the observed
value of $E_R$ ($\sim380$ keV). For completeness, note that carbon can
also be produced by a non-resonant reaction \cite{salpeter}.  This
alternative channel is dominant for sufficiently low temperatures,
$T\simless8\times10^7$ K for canonical values of the parameters used
in \mesa. For the stellar models of this paper, however, the operating 
temperatures for helium burning are larger than this benchmark value. 

\section{Stellar Evolution Simulations} 
\label{sec:evolution}

This section presents results from stellar evolution simulations
carried out using \mesa. Numerical considerations are discussed in
Section \ref{sec:mesa}, along with specification of the initial
conditions and the extent of the parameter space. For comparison,
Section \ref{sec:massevolve} reviews basic trends for the evolution of
massive stars in our universe.  The effects of changing the energy of
the triple alpha resonance are addressed next. Evolutionary tracks in
the H-R diagram and the central density-temperature plane are
described in Section \ref{sec:resevolve}, along with the time
evolution of the nuclear inventory. The yields for carbon, oxygen, and
other alpha elements are then given in Section \ref{sec:yields}.

\subsection{Numerical Considerations and Initial Conditions}
\label{sec:mesa} 

This study uses the computational package \mesa~ to follow the time
evolution of stars \cite{mesaone}. In order to include varying values
for the triple alpha resonance energy, we had to modify the standard
software.  The \mesa~ code controls nuclear reaction rates within a
particular module, which creates a table of integrated cross-sections
for each simulation. Note that these cross sections are the
thermally-averaged quantities $\big\langle\sigma v\big\rangle$.  
Within this module, the cross section for the triple-alpha reaction
$\langle\alpha\alpha\alpha\rangle$ includes the exponential factor
given in equation (\ref{trialpharate}), so that 
\be
\big\langle\alpha\alpha\alpha\big\rangle \propto
\big\langle\alpha\alpha\big\rangle
\big\langle\alpha^{8}\mathrm{Be}\big\rangle \propto 
\exp \left[ -{E_R\over kT}\right] \,, 
\ee 
where $\big\langle\alpha\alpha\big\rangle$ and
$\big\langle\alpha^{8}\mathrm{Be}\big\rangle$ denote the cross
sections for the sub-processes given in equations (\ref{hehe2be}) and
(\ref{hebe2c}), respectively, and $E_R$ denotes the energy of the
triple alpha resonance \cite{nomoto}. We allow the resonance energy to
vary by an increment $\delres$, as defined in equation (\ref{delresdef}).

In addition to variations in the resonance energy for the triple alpha
reaction, as determined by the variable $\delres$, we also explore a
range of stellar masses $M_\ast$ and metallicities $\metal$. The
parameter space for this study thus has three variables and can be 
defined by ${\cal S}=\{\delres,M_\ast,\metal\}$. 

Here we are interested in massive stars that eventually explode as
supernovae. In such stars, the stellar nucleosynthesis continues until
the star produces a degenerate iron core. Since iron is the most
tightly bound nucleus, no further nuclear processing is energetically
favorable, and the star subsequently explodes. In the present
application, we are interested in the final abundances of carbon,
oxygen, and other heavy elements. We evolve the stellar models until
the cores start to produce iron. At this point in evolution, the
abundances of carbon and oxygen become fixed and we can determine
their elemental abundances. For computational convenience, the
simulations are stopped once the core begins to produce iron
(specifically, when the iron mass exceeds 0.1 $M_\odot$). After this
milestone, the time-step in the code becomes increasingly small, and
further evolution becomes computationally expensive.

The stellar mass range is taken to be $M_\ast=15-40 M_\odot$.  The
lower end of the mass range ($15M_\odot$) is chosen for computational
convenience: Since we are interested in massive stars that experience
supernovae, the stellar mass must be larger than the minimum value
$\sim8-10M_\odot$ required to produce an iron core (see \cite{jones}
for a detailed discussion). We choose the slightly larger lower bound
of $15M_\odot$ because the lower mass cutoff can vary with changes in
the input physics (e.g., the value of $\delres$) and because stellar
models near the minimum supernova mass can have difficulty converging.
We also impose an upper mass cutoff ($M_\ast=40M_\odot$). In our
universe, the stellar initial mass function is steep,
$dN/dM_\ast\propto M_\ast^{-2.3}$ \cite{salpeterimf}, so that stars
with masses beyond the cutoff at $40M_\odot$ are rare. Moreover, such
high mass stars are subject to pulsations and other instabilities, and
they experience significant mass loss before exploding as supernovae.
Such complications require specification of a large set of (uncertain)
parameters.  Notice also that the stellar mass distribution could be
different in other universes, so that the proper mass range is not
known.  In any case, we focus on the confined range $M_\ast$ =
$15-40M_\odot$, sampled at integer values of $M_\ast/M_\odot$, to
represent typical stars of high mass. Although this approach does not
fully determine the cosmic abundances, it provides a good description
of how nuclear yields vary with $\delres$.

For the stellar metallicity, we use two values: First, we adopt the
Solar value $\metal=0.02$ because it allows for comparison with nearby
stars in our universe. Since the primordial abundances of heavy
elements are expected to be small, we also consider a much lower
metallicity. Here we adopt the value $\metal=10^{-4}$ characteristic
of the lowest metallicity stars observed in our universe. With such
low metallicity, stellar evolution proceeds similar to the case of
zero metallicity, but the numerical code has better convergence
properties.  With different metallicities, stars have different
starting carbon abundances, which in turn affects the relative
importance of the CNO cycle versus the $p$-$p$ reaction chain. As a
result, stars with different metallicities can enter into their helium
burning phase with different configurations. In practice, however, the
differences are modest, and our results show the same general trends
for both metallicities (see below and the discussions in
\cite{oberhummer,schlattl}).

This paper implicitly assumes that Big Bang Nucleosynthesis (BBN) is
relatively unchanged by variations in the triple alpha reaction. As a
result, the universe emerges from its early epochs with a composition
dominated by hydrogen and helium, and relatively little mass locked up
in heavier isotopes. The early universe does not produce appreciable
amounts of carbon even if the triple alpha reaction rate is much
larger. The universe goes through its hot early phase quickly --- and
with high entropy \cite{kawanocode,walker,pitrou}. By the time the
universe builds up a substantial mass fraction of helium, so that the
triple alpha process has enough raw material to operate, both the
temperature and density are below the values required to produce
substantial amounts of carbon (and below values found in stellar
interiors). This same argument applies to the production of lithium,
which is more easily synthesized than carbon, and has a mass fraction
of only about Li/H $\approx10^{-10}$. For comparison, typical 
abundances for CNO elements are $\sim10^{-16}$, which is five orders 
of magnitude too small to affect Population III stars \cite{pitrou}. 

\subsection{Stellar Evolution for Massive Stars} 
\label{sec:massevolve} 

The general picture for the time development of massive stars is well
known. Here we provide only a brief overview (see \cite{kippenhahn}
for a textbook treatment and \cite{woosley} for a detailed review). 
The majority of time is spent on the main sequence, where stars have
the proper configuration to burn hydrogen into helium. After hydrogen
fuel in the stellar core is exhausted, the star adjusts its structure
and starts to burn helium into carbon through the triple alpha
process. As outlined above, oxygen is also produced through the
reaction (\ref{hec2o}). Changes to the resonance energy affect the
rate of carbon production, and the temperature at which it occurs,
which in turn affects the rate of transforming carbon into
oxygen. After the end of helium burning, the star condenses further
and heats up, so that carbon is burned into neon, primarily through
the reaction $^{12}$C($^{12}$C,$\alpha)^{20}$Ne.  The alpha particles
released through this process are (mostly) absorbed by $^{16}$O nuclei
to make more neon. The relative amounts of carbon and oxygen produced
during the main helium burning phases determine the radial extent of
the star over which these subsequent reactions take place, and how
long they last.

The general evolutionary trend is for the stellar core to become 
both hotter and denser with time, as one nuclear fuel supply is used 
up, and the products from the previous reactions become the fuel for 
the next phase. In the later stages, the core is hot enough for the 
background photons to photo-dissociate $^{20}$Ne into $^{16}$O and 
more alpha particles. Additional alpha elements are then produced, 
especially $^{24}$Mg and $^{32}$S. Another significant process is 
that $^{16}$O nuclei combine to form $^{32}$S, $^{28}$Si, and 
similar nuclei. The eventual formation of a degenerate iron core 
marks the end of this chain of nuclear creation. As this process 
plays out, the carbon and oxygen mass fractions increase during 
the epoch of helium burning, and generally decrease afterwards 
(with some additional production, especially oxygen). In the later 
stages of the evolution, the central region, which becomes the 
iron core, is largely devoid of carbon and oxygen. Moreover, the 
abundances of these nuclei reach a plateau. 

\begin{figure}[tbp]
\centering 
\includegraphics[width=1.0\textwidth,trim=0 150 0 150]{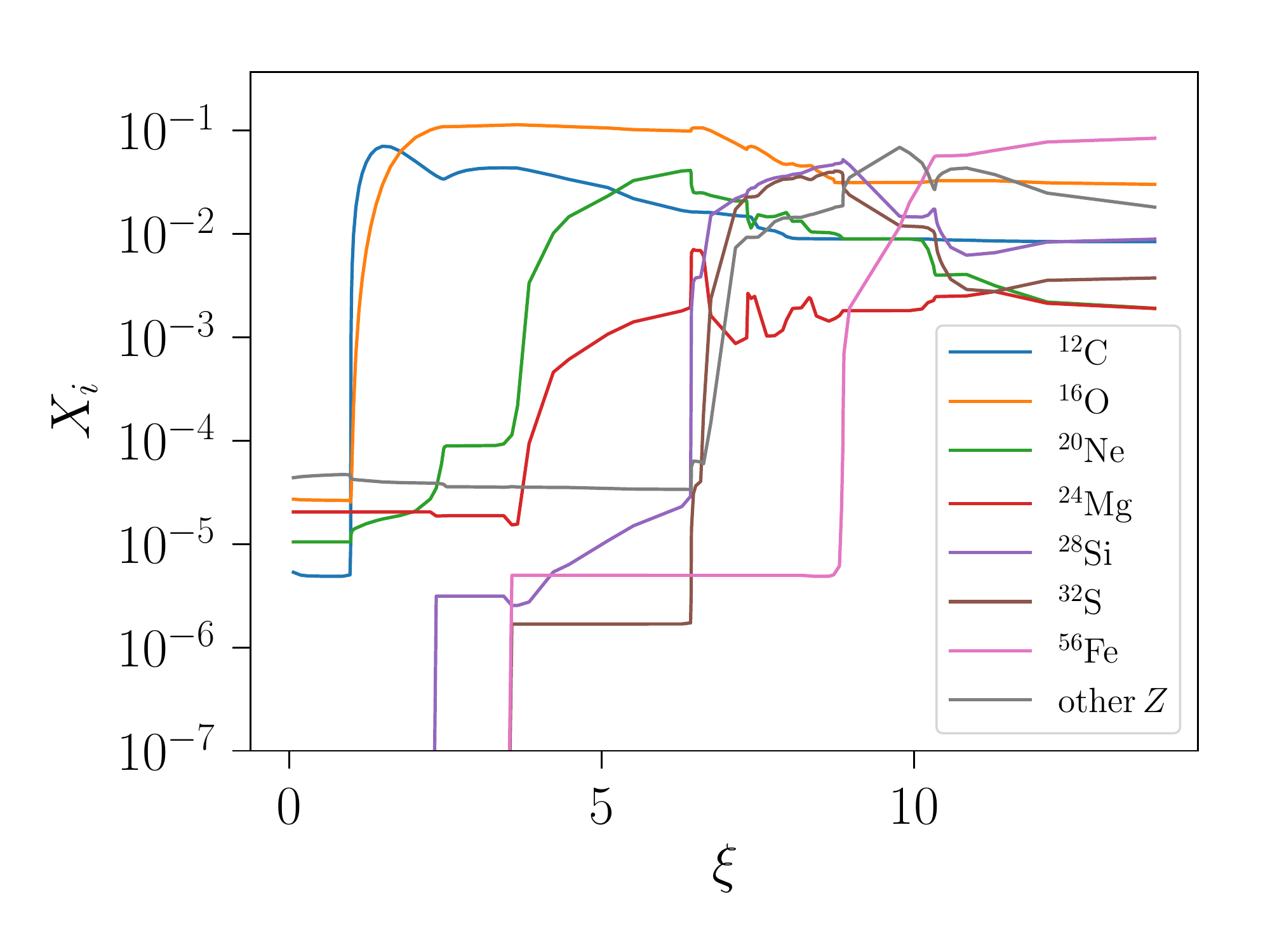}
\vskip 1.20truein 
\caption{Heavy element abundances a function of time for a massive 
star with $M_\ast$ = 15 $M_\odot$, metallicity $\metal=10^{-4}$, 
and the triple alpha resonance properties of our universe. The yields
are given as mass fractions $X_i$. The time variable is a logarithmic
measure of the time remaining before the end of the simulation at
$\tend$ when an iron core develops. As time elapses, progressively
heavier nuclei are synthesized and their abundances grow, including
carbon (blue), oxygen (orange), neon (green), magnesium (red), silicon
(purple), sulfur (brown), iron (pink), and all other metals (gray).
The abundances of the alpha elements increase, reach a maximum, and
subsequently decline back down to an essentially constant value. }
\label{fig:xvtime} 
\end{figure} 

\begin{figure}[tbp]
\centering 
\includegraphics[width=1.0\textwidth,trim=0 150 0 150]{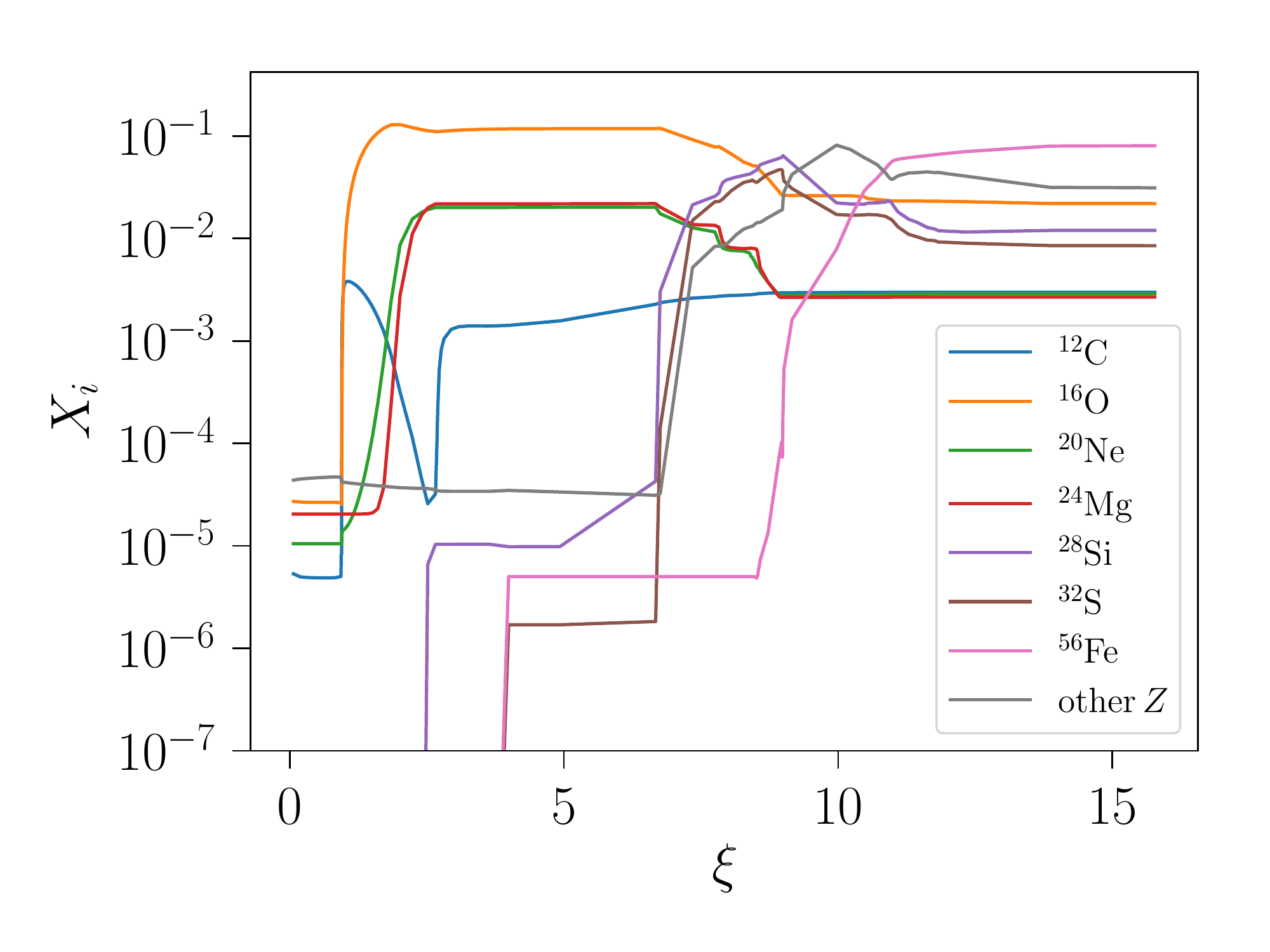}
\vskip 1.20truein 
\caption{Heavy element abundance a function of time for a massive 
star with $M_\ast$ = 15 $M_\odot$, metallicity $\metal=10^{-4}$, 
and resonance increment $\delres$ = 100 keV. The yields are given as
mass fractions $X_i$.  The time variable is a logarithmic measure of
the time remaining before the end of the simulation at $\tend$ when an
iron core develops. As time elapses, progressively heavier nuclei are
synthesized and their abundances grow, including carbon (blue), oxygen
(orange), neon (green), magnesium (red), silicon (purple), sulfur
(brown), iron (pink), and all other metals (gray).  The abundances of
the alpha elements increase, reach a maximum, and subsequently decline
back down to an essentially constant value. }
\label{fig:xvtime100} 
\end{figure} 

This behavior is illustrated by Figures \ref{fig:xvtime} and
\ref{fig:xvtime100}, which show the mass fractions of the most
important nuclei as a function of time for stellar models with mass
$M_\ast$ = 15 $M_\odot$ and resonance increment $\delres$ = 0 and +100
keV, respectively. Because the time for each subsequent nuclear
burning stage is shorter than the previous one, we present the results
in terms of the time variable 
\be
\xi \equiv - \log_{10} \left[ 1 - {t \over \tend} \right]\,,
\ee
where the time $\tend$ marks the formation of the iron core and end of
the numerical simulation. The variable $\xi$ thus provides a
logarithmic measure of the time remaining before the star explodes.
Note that the variable $\xi\to\infty$ in the limit $t\to\tend$. For 
the curves shown in Figures \ref{fig:xvtime} and \ref{fig:xvtime100}, 
the value of $\xi_n$ at the final ($n$th) time-step is plotted as
$\xi_n=\xi_{n-1}+1$.

As time increases, the mass fractions $X_i$ of progressively heavier
elements start to grow. The fraction of mass contained in carbon
(blue), oxygen (orange), neon (green), magnesium (red), and silicon
(purple) all increase, reach a maximum value, and then decrease down
to an essentially constant value. Near the end of its life, the star
produces a substantial amount of sulfur (brown), which is then
transformed into iron (pink). Significantly, by the end stages of
evolution when iron starts to accumulate in the core
($\xi\simgreat10$), the mass fractions for all of the alpha elements
are nearly flat/constant. As a result, the carbon and oxygen yields
reported in this paper correspond to the abundances of the star at the
time when the iron core develops.

The stellar evolution code (\mesa) does not follow the supernova
explosion. However, the stellar core itself contains little
carbon/oxygen at this point, so that these elements are not lost to
the neutron star remnant that forms out of the iron core. We also
assume that the supernova explosion successfully detonates, so that no
carbon and oxygen in the outer stellar layers are lost due to fallback
onto the remnant. On the other hand, some additional nuclear
processing does take place in the supernova ejecta, but these
modifications to the carbon and oxygen inventories are beyond the
scope of this present treatment. Earlier work argues that the carbon
abundances are not altered appreciably by the supernova shock wave
\cite{schlattl}, whereas the abundance of oxygen can be depleted by
$\sim10\%$. The carbon yields are determined to higher accuracy.

\subsection{Stellar Evolution with Varying Carbon Resonance Energy}  
\label{sec:resevolve} 

Variations in the energy of the triple alpha resonance have important
implications for the final yields of carbon and oxygen produced by
massive stars, but relatively modest effects on the overall trajectory
of stellar evolution. Before determining the carbon and oxygen yields
as function of $\delres$ (see Section \ref{sec:yields}), we consider
how changes in the resonance energy affect the evolution of massive 
stars. All of the stars in the mass range of interest tend to exhibit 
qualitatively similar behavior, so this discussion includes only 
a few representative examples. 

\begin{figure}[tbp]
\centering 
\includegraphics[width=1.0\textwidth,trim=0 150 0 150]{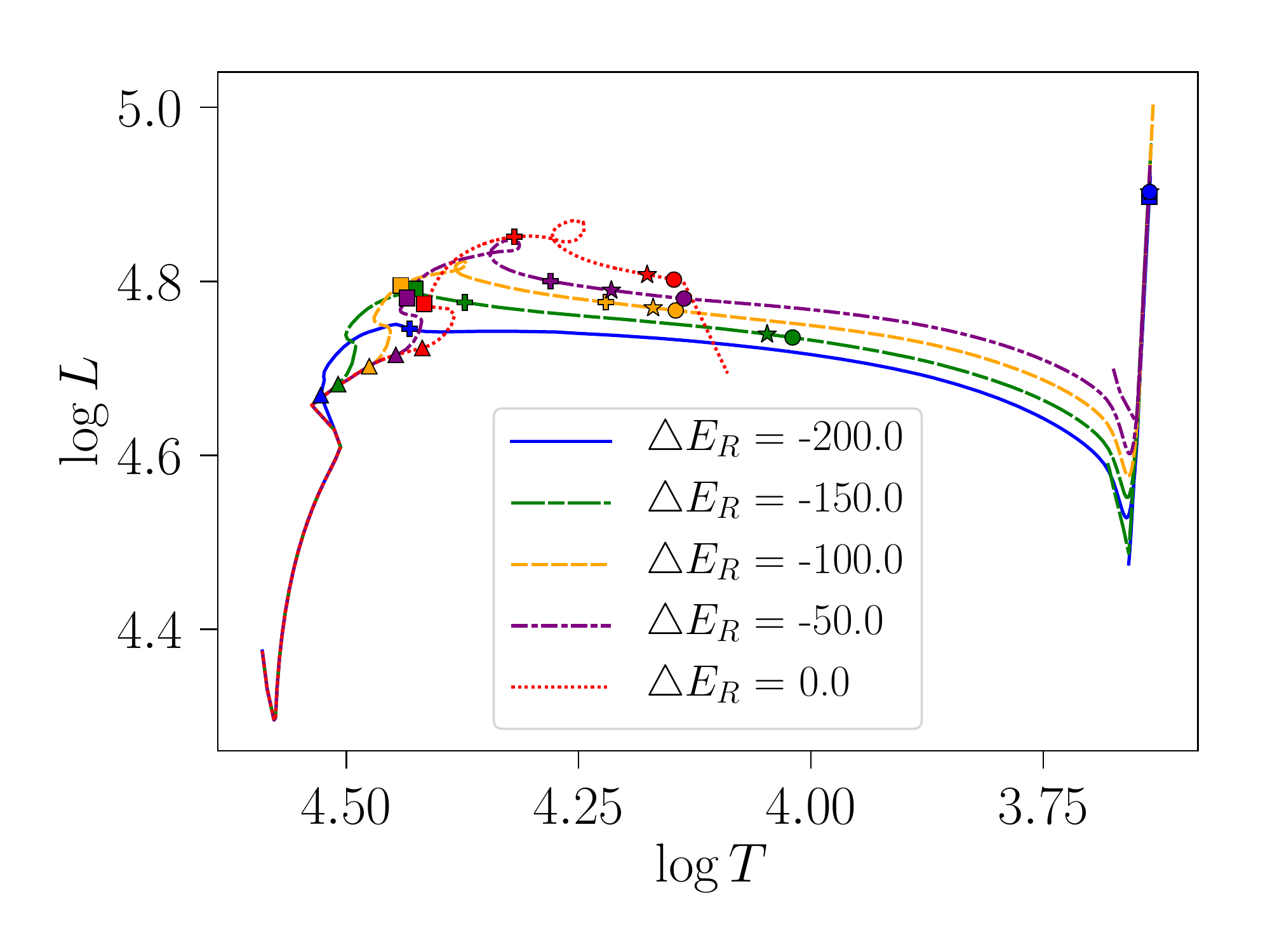}
\vskip 1.20truein 
\caption{H-R diagram for the evolution of a star with mass $M_\ast$ = 
15 $M_\odot$ and metallicity $\metal=10^{-4}$. Tracks for shown for
negative increments of the triple alpha resonance energy $\delres$ =
--200 keV (blue), --150 keV (green), --100 keV (orange), --50 keV 
(violet), and 0 (red). The symbols on the tracks mark the points 
where the alpha elements start to be produced, including carbon
(triangles), oxygen (squares), neon (plus signs), magnesium (stars),
and silicon (circles). }
\label{fig:hr15negative} 
\end{figure} 

\newpage 

\begin{figure}[tbp]
\centering 
\includegraphics[width=1.0\textwidth,trim=0 150 0 150]{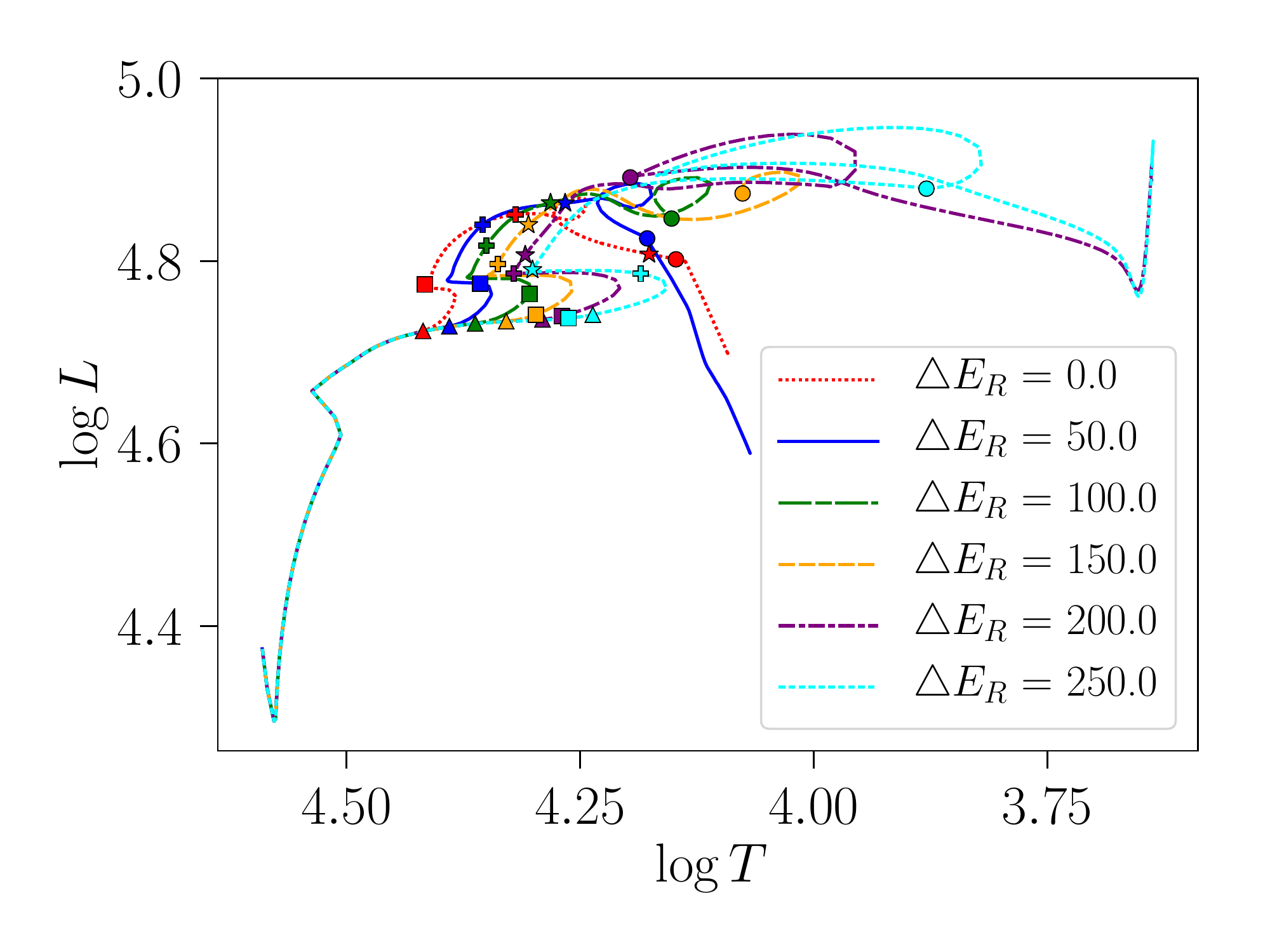}
\vskip 1.20truein 
\caption{H-R diagram for the evolution of a star with mass $M_\ast$ = 
15 $M_\odot$ and metallicity $\metal=10^{-4}$. Tracks for shown for
positive increments of the triple alpha resonance energy $\delres$ =
0 (red), 50 keV (blue), 100 keV (green), 150 keV (orange), 200 keV 
(violet), and 250 keV (cyan). The symbols on the tracks mark the points 
where the alpha elements start to be produced, including carbon
(triangles), oxygen (squares), neon (plus signs), magnesium (stars),
and silicon (circles). }
\label{fig:hr15positive} 
\end{figure} 

The Hertzsprung-Russell (H-R) diagram is shown in Figures
\ref{fig:hr15negative} and \ref{fig:hr15positive} for massive stars
with $M_\ast$ = 15 $M_\odot$ and metallicity $\metal=10^{-4}$.  The
two figures show evolutionary tracks for stars with varying triple
alpha resonance energies, corresponding to the range $\delres$ = --200
keV to +250 keV. For all of the stars, evolutionary tracks start on
the zero-age main-sequence, on the left side of the diagram.  The
stars then move upward in the diagram as their hydrogen fuel becomes
depleted.  The symbols on the tracks mark the points where alpha
elements of interest start to be produced, including carbon
(triangles), oxygen (squares), neon (plus signs), magnesium (stars),
and silicon (circles). The symbols denote the points where the mass
contained in these elements doubles from its initial value. Note that
the tracks start to diverge with the onset of helium burning (carbon
production). The models with the lowest values of $\delres$ branch 
off first, and the tracks continue to branch in order of increasing
$\delres$, corresponding to higher temperatures required for the
triple alpha reaction to operate. At late times, the tracks tend to
converge, although the stellar models with $\delres\sim0$ have lower
luminosity at the endpoint (given by the formation of a degenerate
core).  Nonetheless, all of stars have similar internal configurations
at the end of their tracks as determined by their central temperatures
and densities (see Figure \ref{fig:center} below). 

At intermediate times, the tracks in the H-R diagram show more
complicated behavior. In Figure \ref{fig:hr15negative}, the tracks for
stellar models with $\delres=-200$, --150, and --100 keV execute
additional loops compared to the others, i.e., the tracks go back and
forth in the H-R diagram. Additional loops are also seen in Figure
\ref{fig:hr15positive} for stellar models with large $\delres$ = 150
-- 250 keV.  Such loops are common signatures of the later stages of
nuclear burning in massive stars, and are extremely sensitive to the
input physics (see \cite{kippenhahn} for an in-depth discussion).
Under the standard ordering, the production of nuclei proceeds in
order of increasingly atomic number: carbon, oxygen, neon, magnesium,
and then silicon. The symbols in Figures \ref{fig:hr15negative} and
\ref{fig:hr15positive} generally follow this pattern, but exceptions
arise for extreme values of $\delres$. For example, for $\delres$ =
--200 keV (the most negative value shown in Figure
\ref{fig:hr15negative}), oxygen production (marked by the blue square)
is delayed. For $\delres$ large and negative, the temperature for the
triple alpha reaction is low, and little oxygen is produced. The
carbon produced is then processed into neon (plus symbol), with oxygen
produced much later during shell burning in the later stages of
evolution. As another example, for $\delres$ = +250 keV (Figure
\ref{fig:hr15positive}), the temperature for the triple alpha process
is so high that oxygen production (cyan square) occurs before carbon
(cyan triangle).

Another way to illustrate stellar evolution is to plot the central
temperature of the star versus the central density as the star
evolves. The general trend is for stellar cores to become both hotter
and denser as they burn through one nuclear fuel source and move on to
the next heavier one. Figure \ref{fig:center} illustrates this type of
behavior for massive stars with $M_\ast=30M_\odot$ and low metallicity
$\metal=10^{-4}$. Tracks are shown for a range of triple alpha
resonance energies corresponding to $\delres$ = --200 (blue), --100
(green), 0 (red), 100 (gold), 200 (purple), and 300 keV (cyan). The
evolution of the central stellar conditions in this diagram displays
the same basic evolutionary behavior for all values of the resonance
energy. For the tracks shown here, the stars start on the zero-age
main-sequence. Stellar cores reside in the lower left portion of the
diagram over most of their lifetimes as they burn hydrogen into
helium, with central temperature $T_{\rm c}\sim3\times10^7$ K. The
stars then move toward the upper right as they produce ever larger
nuclei. In our universe, stellar cores typically have temperature
$T_{\rm c}\sim2\times10^8$ K and density $\rho_{\rm c}\sim10^3$ g
cm$^{-3}$ during their helium burning phases \cite{schaller}, when the
triple alpha process is active. This epoch is marked by the triangles
in the figure, which shows how the central temperature for carbon
production becomes progressively hotter as $\delres$ increases. The
density increases as well, so that the stars stay on nearly the same
trajectory in the diagram. As a result, the curves shown in Figure
\ref{fig:center}, with varying values of $\delres$, show relatively
little spread as the stars sequentially produce helium, carbon,
oxygen, and then neon.  The stars evolve until they reach the upper
right portion of the diagram, as they eventually develop degenerate
iron cores.  Superimposed on this general evolutionary trend, the
tracks in the diagram show minor excursions from simple monotonic
behavior, primarily during the later stages of nuclear burning.

\begin{figure}[tbp]
\centering 
\includegraphics[width=1.0\textwidth,trim=0 150 0 150]{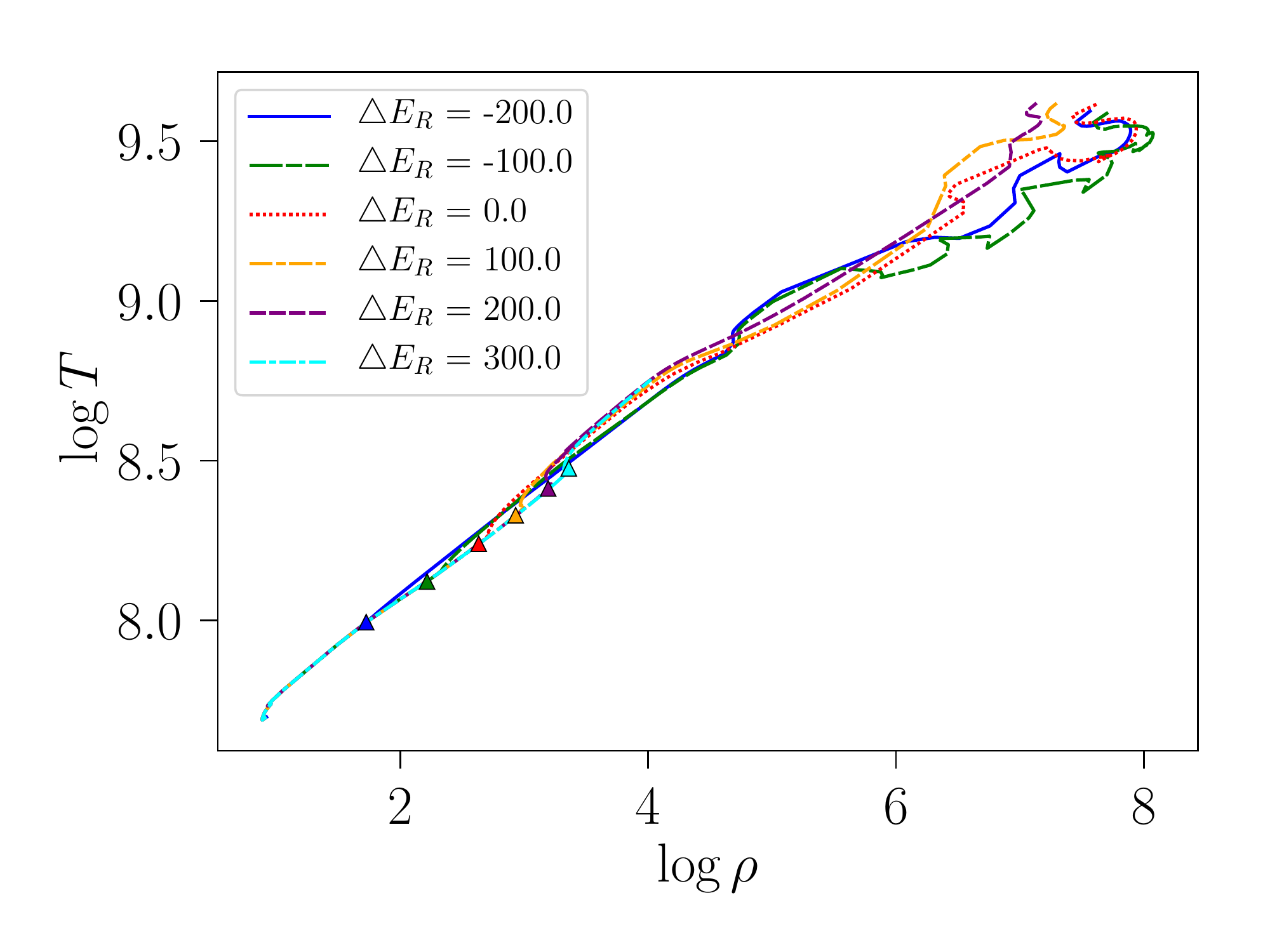}
\vskip 1.20truein 
\caption{Central density (g cm$^{-3}$) and central temperature (K) of
a massive star over its lifetime. The stellar mass $M_\ast$ = 30 
$M_\odot$ and the initial metallicity $\metal=10^{-4}$. 
The curves shows the results obtained from stellar evolution
simulations using different resonance energies for the triple alpha
reaction corresponding to $\delres=-200\to+300$ keV. Stars start in 
the lower left part of the diagram with the right configuration to
burn hydrogen and end up in the upper right part of the diagram when
they develop degenerate iron cores. The triangles mark the points 
where helium burning through the triple alpha process begins. }
\label{fig:center} 
\end{figure}   

The behavior depicted in Figure \ref{fig:center} can be understood in
terms of the stellar structure equations \cite{clayton,hansen}.  In
order for the star be be in hydrostatic equilibrium, the central
pressure is given by
\be
P_{\rm c} \approx \left({\pi \over 36}\right)^{1/3} G M_\ast^{2/3}  
\rho_{\rm c}^{4/3} \,. 
\label{centpress} 
\ee
While the star remains operational, its pressure support is dominated
by the ideal gas law, so that the central pressure is also given by
\be
P_{\rm c} \approx {1 \over \langle{m}\rangle} \rho_{\rm c} kT_{\rm c} \,,
\label{idealgas} 
\ee
where $\langle{m}\rangle$ is the mean mass of the ions. These two 
considerations imply a temperature-density relation of the form 
\be
kT_{\rm c} = \left({\pi\over36}\right)^{1/3} 
\langle{m}\rangle G M_\ast^{2/3} \rho_{\rm c}^{1/3} \,. 
\label{rhotemp} 
\ee
This result provides an approximate description of the curves shown in
Figure \ref{fig:center}. The numerical results are slightly less steep
than $T_{\rm c}\sim\rho_{\rm c}^{1/3}$, primarily due to the inclusion
of radiation pressure at high temperatures. The nuclear composition
determines the mean ionic mass $\langle{m}\rangle$, which also changes
the slope of the relation and produces the departures from monotonic 
behavior, as shown in the diagram. 

\subsection{Yields for Carbon, Oxygen, and Larger Nuclei} 
\label{sec:yields} 

\begin{figure}[tbp]
\centering 
\includegraphics[width=1.0\textwidth,trim=0 150 0 150]{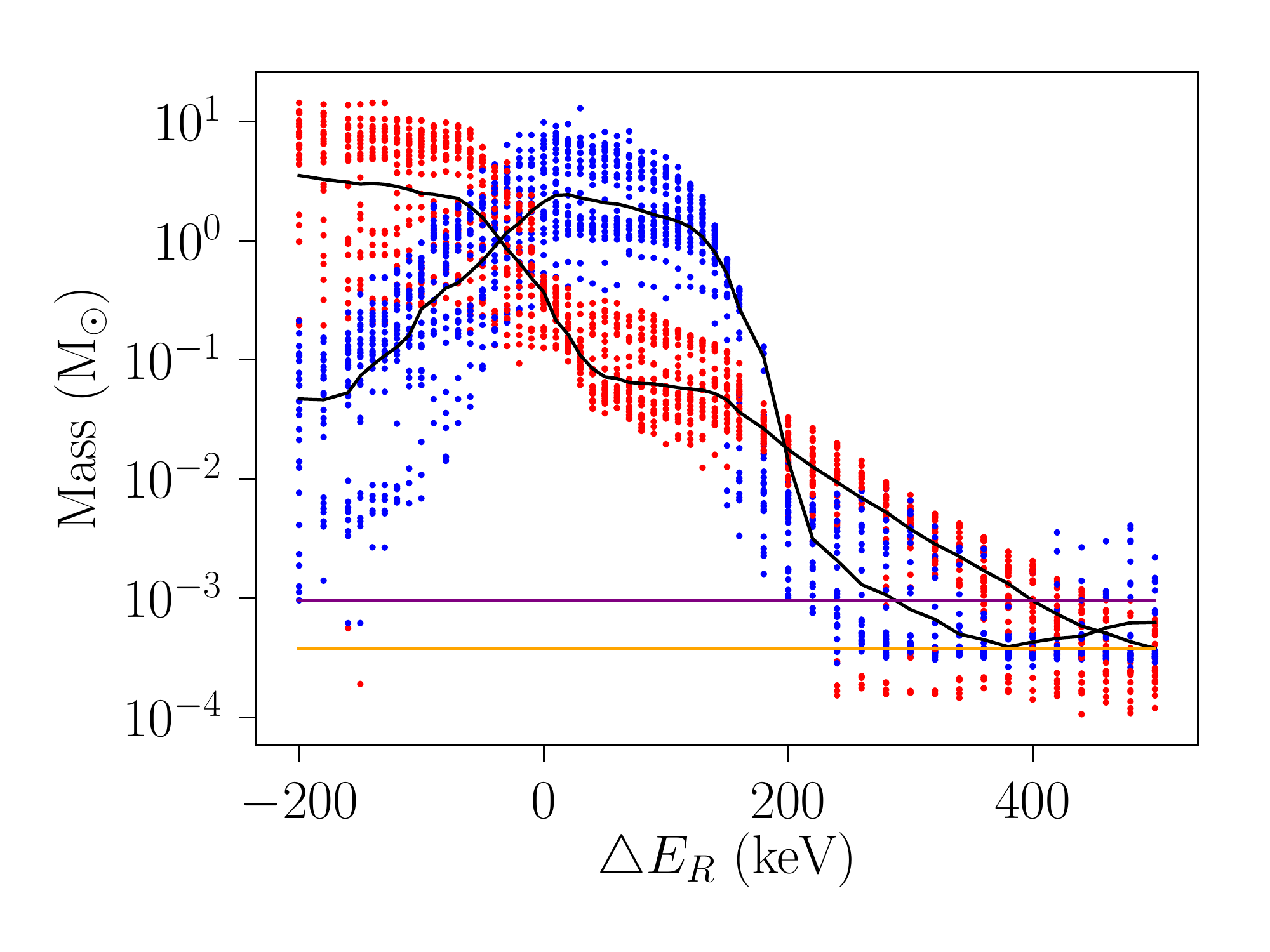}
\vskip 1.20truein
\caption{Carbon and oxygen production yields in massive stars as a
function of the $0^+$ resonance energy of the carbon-12 nucleus 
(for initial metallicity $\metal=10^{-4}$). The resonance energy is
specified on the horizontal axis by the difference $\delres$ from the
value in our universe (in keV). Results are shown for stellar
evolution simulations with stellar masses in the range $M_\ast$ = 
$15-40M_\odot$.  The red circles show the resulting yields for carbon
(in $M_\odot$), whereas the blue circles show the corresponding yields
for oxygen. The heavy black curves show the expectation value (in
$M_\odot$ per star) obtained with a weighted average over the stellar
initial mass function in the specified range. The horizontal lines
show the expectation values for the starting abundances of carbon
(lower) and oxygen (upper) corresponding to $\metal=10^{-4}$.  The
yields for carbon and oxygen fall below their starting values for 
$\delres\approx+480$ keV and $\delres\approx+250$ keV, 
respectively. } 
\label{fig:COfractionzero} 
\end{figure}

\begin{figure}[tbp]
\centering 
\includegraphics[width=1.0\textwidth,trim=0 150 0 150]{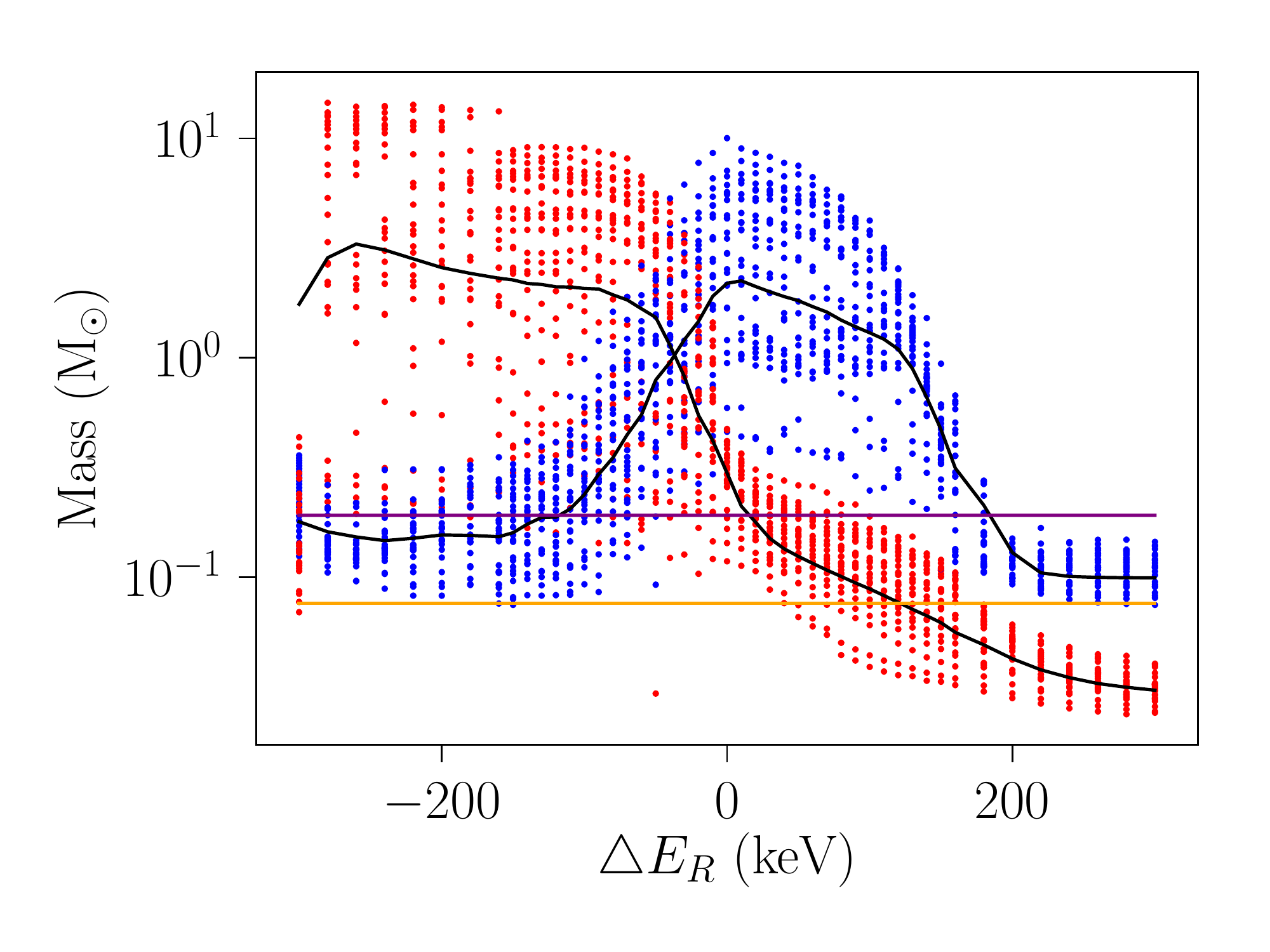}
\vskip 1.20truein
\caption{Carbon and oxygen yields in massive stars as a function 
of the $0^+$ resonance energy of the carbon-12 nucleus (for solar 
metallicity). The resonance energy is specified by the difference
$\delres$ from the value in our universe (in keV). Results are shown
for stellar evolution simulations with stellar masses in the range
$M_\ast=15-40M_\odot$.  The red circles show the resulting yields for
carbon (in $M_\odot$), whereas the blue circles show the corresponding
yields for oxygen. The heavy black curves show the expectation value
(in $M_\odot$ per star) obtained with a weighted average over the
stellar initial mass function in the specified range. The horizontal
lines show the expectation values for the starting abundances of
carbon (lower) and oxygen (upper) corresponding to $\metal$ = 
$\metal_\odot$. The yields fall below the starting values for
$\delres\approx+180$ keV for oxygen and $\delres\approx+120$ keV
for carbon. }
\label{fig:COfractionzsun} 
\end{figure}

We have carried numerical simulations for stellar models corresponding
to the parameter space outlined in Section \ref{sec:mesa}. The
resulting carbon and oxygen yields are shown as a function of the
change $\delres$ in the triple alpha energy level in Figure
\ref{fig:COfractionzero} (for low metallicity $\metal=10^{-4}$) and
Figure \ref{fig:COfractionzsun} (for solar metallicity $\metal$ =
$\metal_\odot$).  In these figures, the mass in carbon (blue symbols)
and oxygen (red symbols) are plotted for each stellar evolution model
(specified by stellar mass $M_\ast$ and resonance energy increment
$\delres$). The stellar models are run until the stars begin to
produce iron cores, at which time no further processing of carbon and
oxygen occurs.  The expectation value of the yields (in units of
$M_\odot$ per star) is shown as the thick black curves, which were
obtained by integrating over the range of stellar masses, weighted by
the stellar initial mass function. The curves were also smoothed by 
averaging over adjacent bins. The horizontal lines in the figures 
show the expectation values for carbon (lower gold line) and oxygen 
(upper purple line) appropriate for the starting metallicity.

Figures \ref{fig:COfractionzero} and \ref{fig:COfractionzsun} show a
number of interesting trends: When the resonance energy is lowered, so
that $\delres<0$, the carbon yields are {\it higher} than those of our
universe. Our numerical simulations are only carried out over the full
mass range for the values of $\delres$ shown, but the carbon yields
are larger than those in our universe for all $\delres<0$. Moreover,
although this finding is consistent with previous results
\cite{oberhummer,schlattl}, the fact that lower resonance levels lead
to more carbon production is not widely appreciated. On the other
hand, the oxygen abundances decrease for lower values of $\delres$. 
For both metallicities under consideration, the expectation values for
carbon and oxygen yields are equal for $\delres\approx-35$ keV. Notice 
also that the resonance level cannot be made arbitrarily low: If the 
resonance is lowered by more than $\sim380$ keV, the reaction is no
longer energetically favorable (see equation [\ref{resenergy}]). In
practice, the resonance energy must be moderately higher in order for
the star to remain stable (see equation [\ref{resconstraint}]), so
that the lower limit becomes $\delres\simgreat-300$ keV.

The carbon yields decrease with increasing values of $\delres$, as
expected. For higher resonance energies, the temperature required for
the triple alpha reaction to operate increases, and much of the carbon
produced can be immediately burned into oxygen.  For low metallicity
stars (Figure \ref{fig:COfractionzero}), the carbon yields steadily
decrease with increasing $\delres$ and cross the starting values at
$\delres\approx+500$ keV. The oxygen yields increase with moderate
increases in the resonance energy. For $\delres\simgreat130-150$ keV,
however, the oxygen yields steadily decrease with further increases in
the resonance energy. The oxygen abundances fall below that of carbon
for $\delres\sim200$ keV, and become less than the starting value for
$\delres\sim300$ keV. To summarize, stars in other universes can 
support carbon production over the range in resonance energy 
increment given by 
\be 
-300~{\rm keV} \simless\delres\simless 500~{\rm keV}\,. 
\qquad {\rm (carbon)} 
\label{resrangecar} 
\ee
A moderately smaller range in $\delres$ allows for oxygen production, 
\be 
-300~{\rm keV} \simless\delres\simless 300~{\rm keV}\,. 
\qquad {\rm (oxygen)} 
\label{resrangeoxy} 
\ee
The total ranges in $\delres$ are thus $\sim800$ keV for carbon 
and $\sim600$ keV for oxygen. 

For solar metallicity stars $\metal=\metal_\odot$ (Figure
\ref{fig:COfractionzsun}), the ranges in resonance energy for which
stars produce more carbon and oxygen than their starting values are
somewhat smaller than for low metallicity.  Carbon production exceeds
the initial supply for all negative $\delres$ and for positive
resonance energy increments up to $\delres\approx$ +160 keV. Oxygen
production exceeds the starting value for resonance increments in the
range --150 keV $\simless\delres\simless$ +200 keV.

\begin{figure}[tbp]
\centering 
\includegraphics[width=1.0\textwidth,trim=0 150 0 150]{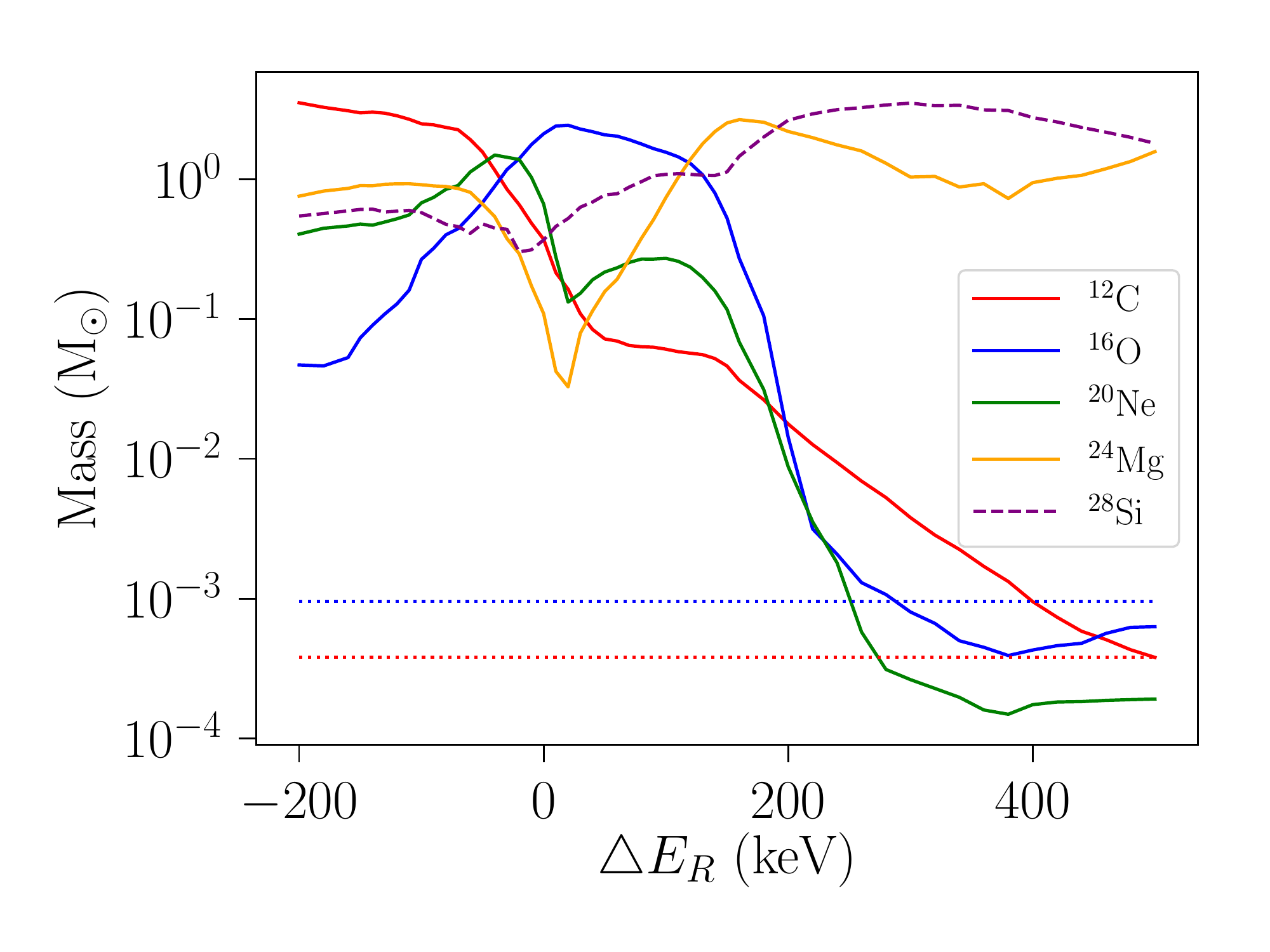}
\vskip1.20truein 
\caption{Yields from massive stars for an ensemble of alpha elements 
as a function of the triple alpha resonance energy (specified by the 
energy increment $\delres$).  These simulations have starting
metallicity $\metal$ = $10^{-4}$. Each curve shows the weighted-mean
mass per star for a given alpha element. Yields are shown here for
carbon (red), oxygen (blue), neon (green), magnesium (orange), and
silicon (dashed purple). Changes to the resonance energy $\delres$ are
given in keV, and the expectation values for the yields are given in
$M_\odot$. }
\label{fig:allzero} 
\end{figure} 

\newpage

\begin{figure}[tbp]
\centering 
\includegraphics[width=1.0\textwidth,trim=0 150 0 150]{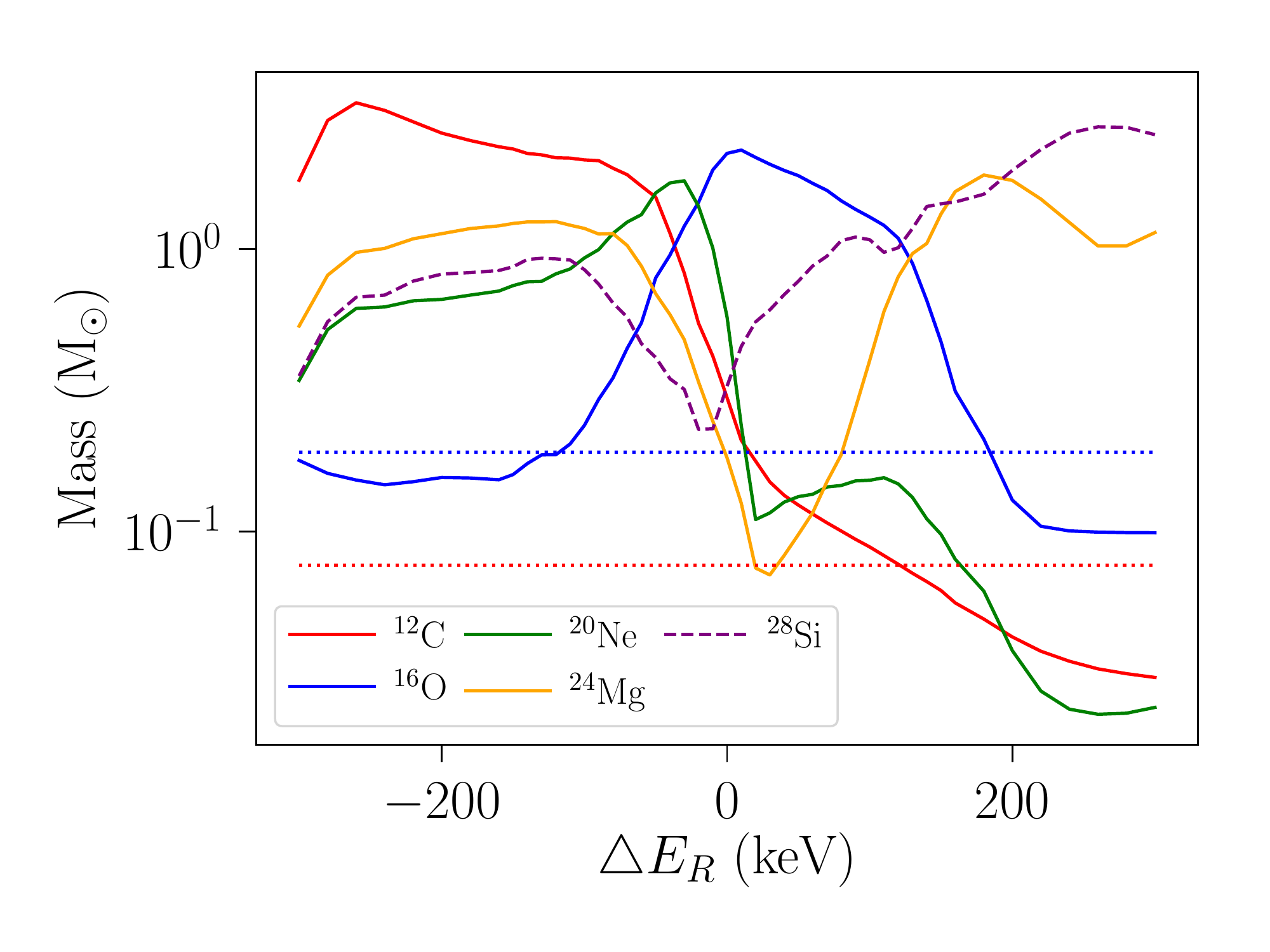}
\vskip1.20truein 
\caption{Yields from massive stars for an ensemble of alpha elements 
as a function of the triple alpha resonance energy (specified by the
energy increment $\delres$). These simulations have starting
metallicity $\metal=\metal_\odot$. Each curve shows the weighted-mean
mass per star for a given alpha element. Yields are shown here for
carbon (red), oxygen (blue), neon (green), magnesium (orange), and
silicon (dashed purple). Changes to the resonance energy $\delres$ are
given in keV, and the expectation values for the yields are given in
$M_\odot$. }
\label{fig:allzsun} 
\end{figure} 

The nuclear abundances for a wider range of elements synthesized in
massive stars are shown in Figures \ref{fig:allzero} and
\ref{fig:allzsun} for initial metallicities $\metal=10^{-4}$ and
$\metal=\metal_\odot$, respectively. For each star, the numerical
results determine the mass locked up in each element at the end of the
simulation. These masses are then averaged over stellar mass, for the
range $M_\ast=15-40M_\odot$, weighted by the stellar initial mass
function. The functions are then smoothed over adjacent bins. The
resulting expectation value (in $M_\odot$ per star) for each element
is shown as a function of the change $\delres$ in the triple alpha
resonance energy for carbon (red), oxygen (blue), neon (green),
magnesium (orange), and silicon (purple). These simulations are run
until the stars begin to develop iron cores. At this juncture, the
abundances of the first four of these elements have reached their
asymptotic values, but some of the silicon could still be processed
into iron at later times.  As a result, the yields for silicon are
more uncertain than those of the other elements, so that these
estimates are depicted by dashed curves. The lower horizontal lines in
Figures \ref{fig:allzero} and \ref{fig:allzsun} show the expectation
values for carbon (lower red) and oxygen (upper blue) at the start of
the simulations. As outlined above, they determine the ranges in
$\delres$ for which stars can produce carbon and oxygen (see equations
[\ref{resrangecar}] and [\ref{resrangeoxy}]).

The results depicted in Figures \ref{fig:allzero} and
\ref{fig:allzsun} show the same basic trends: As noted above, the
carbon yields steadily decrease as the triple alpha energy increment
$\delres$ increases. In contrast, the oxygen yields increase with
$\delres$ until the resonance energy is higher than that of our
universe, then saturate as they reach a broad peak, and decrease with
further increases in the resonance energy. The neon yields are
primarily decreasing functions of the resonance energy, but show more
structure than the results for carbon.  In particular, both neon
curves have a local maximum for a small negative value of $\delres$,
and another local maximum near $\delres\approx100$ keV.  To leading
order, the magnesium yields are anti-correlated with those of oxygen,
with a minimum near the resonance level of our universe, and larger
values for both positive and negative values of the resonance
increment $\delres$.  Silicon yields generally grow with increasing
$\delres$, especially for large and positive values, but show a
moderate deficit near $\delres\sim0$. All of these trends are present
in the results for both metallicities, although the detailed shape of
the abundance curves vary (notice also that the axes are different in
Figures \ref{fig:allzero} and \ref{fig:allzsun}).

To a rough approximation, the abundances of these alpha elements
exhibit zero-sum behavior, with their sum nearly constant across the
range in $\delres$ of interest. Each of the five species is the most
abundant element (by mass) for a small range of $\delres$.  As the
resonance energy increases, the isotope that has the peak abundance
varies from carbon to neon, oxygen, magnesium, and then silicon. With
the exception of neon --- and the neon peak is also the least
prominent --- this ordering follows that of increasing atomic number.
Under typical stellar conditions, however, the temperature required
for neon burning is slightly lower than that of oxygen burning, so the
usual `onion-skin' structure of high mass stars also follows this
ordering \cite{clayton,kippenhahn}. The basic trend shown in Figures
\ref{fig:allzero} and \ref{fig:allzsun} is thus expected: As $\delres$
increases, the triple alpha process requires a higher central
temperature, which favors the production of heavier nuclei.

\section{Implications and Interpretation} 
\label{sec:context} 

Results from the previous section show that the resonance energy for
the triple alpha reaction can vary over a range of $\sim800$ keV and
still allow carbon to be produced. With this viable range specified,
this section discusses how much carbon is necessary for habitability,
alternate paths for carbon production via stable beryllium-8, and
implications for the fundamental constants of physics.

\subsection{Observed Carbon Abundances} 
\label{sec:obscarbon} 

With the value observed for the triple alpha resonance energy in our
universe (corresponding to $\delres=0$), stellar nucleosynthesis can
account for the observed carbon abundances.  As shown in the previous
section, massive stars produce carbon-to-oxygen ratios [C/O] of order
unity. This finding is consistent with the value found for the Galaxy
as a whole, [C/O] $\approx$ 0.67, and that for the Solar System [C/O]
$\approx$ 0.51 \cite{cameron,lodders}. The observed range of
carbon-to-oxygen ratios varies from region to region, with [C/O]
$\approx0.25-0.75$.  Although observed [C/O] values are consistent
with predictions from stellar evolution calculations for massive
stars, the carbon inventory of the Galaxy also gets substantial
contributions from intermediate mass stars.

In contrast to the [C/O] ratios quoted above, the observed value for
Earth is substantially lower, with estimates falling in the range
[C/O] $\approx0.002-0.01$ \cite{allegre,marty}. As a result, the
carbon abundance of Earth is about two orders of magnitude lower than
that of the Solar System, the Galaxy, and the universe as a whole.
The minimum carbon abundance for habitability remains unknown at the
present time. Nonetheless, given that Earth is the only astrophysical
environment known to support life, and that its carbon supply is
highly depleted, this carbon threshold could be as much as $\sim100$
times lower than the observed cosmic abundance. From the results shown
in Figure \ref{fig:COfractionzero}, stars in other universes could
produce carbon at Earth-like levels for triple alpha resonance
energies corresponding to increments as large as $\delres\sim+300$
keV.

On the other hand, the estimated [C/O] ratio for Venus is comparable
to that of Earth and hence also sub-solar \cite{morgananders}. One
interpretation of this finding is that rocky planets are ineffective
at capturing and holding onto carbon, so that the formation of
terrestrial planets generically results in [C/O] ratios that are
similarly low. However, the formation location is also important:
C-type (carbonaceous) asteroids, which represent the majority of these
rocky bodies, contain a large abundance of carbon \cite{norton}, with
chemical compositions comparable to that of the Sun (after accounting
for the depletion of hydrogen, helium, and other volatiles). Such
asteroids are thought to be left-over building blocks from the planet
formation process, and could in principle build planets with higher
carbon inventories than Earth. With the wide range of carbon
abundances inferred for bodies in our Solar System, and the diversity
of worlds now being discovered around other stars, the expected carbon
abundance for planets in systems with a given stellar composition
remains to be determined \cite{raymond}.  For completeness, we also
note that low [C/O] ratios could in principle be produced through the
enhancement of oxygen on terrestrial planets.

\subsection{The Possibility of Stable Beryllium-8} 
\label{sec:stablebe8} 

The above discussion indicates that the energy level of the carbon
resonance can vary over a range of $\sim800$ keV while allowing
massive stars to synthesize carbon at acceptable levels. Much larger
variations could render the universe sterile, by making carbon
energetically disfavored for large negative $\delres$ (see equation
[\ref{resenergy}]), and by shutting down carbon production for large
positive $\delres$. For comparison, the $^8$Be nucleus fails to have a
bound state by only 92 keV. In order words, the changes to nuclear
binding energies required to render carbon production untenable is
much larger than the changes necessary for the universe to have stable
$A=8$ nuclei.

With the existence of a stable isotope with mass number $A=8$,
universes no longer need the triple alpha process to produce carbon
\cite{agalpha,higa}. Helium burning can proceed through the arguably
more natural reaction of equation (\ref{hehe2be}), which would
primarily proceed in the forward direction with stable $^8$Be as the
end product. The beryllium could then be processed into carbon through
equation (\ref{hebe2c}) without the need for an enhanced reaction
rate.  The subsequent carbon production could take place in later
evolutionary states of the same star, or much later in future stellar
generations.

In order for a stable $^8$Be nuclei to exist, the required changes to
the fundamental constants are only about 1 -- 2\%.  The size of these
variations can be estimated by detailed calculations using Lattice
Chiral Effective Field Theory \cite{epelbaum}, by simpler models of
the nucleon-nucleon potential \cite{ekstrom}, and/or by
straightforward analytic arguments \cite{agalpha,davies}. Moreover,
the necessary changes to the strengths of the nuclear and
electromagnetic forces are roughly comparable. If the fundamental
forces have different strengths than in our universe, then both the
binding energy of $^8$Be and that of the constituent $^4$He nuclei
will change. Significantly, the aforementioned effective field theory
calculations show that the binding energies of $^8$Be and its building
blocks do not change at the same rate as the fundamental constants are
varied \cite{epelbaum,epelbaum2011}. This mismatch is necessary for
$^8$Be to become stable: If the derivative of the $^4$He binding
energy with respect a fundamental parameter is equal to half the
derivative of the $^8$Be binding energy with respect to the same
parameter, then both sides of equation (\ref{hehe2be}) would scale
together, and the forward reaction would not be energetically favored.

\subsection{Triple Alpha Resonance and the Fundamental Constants} 
\label{sec:funconst} 

The focus of this paper is to explore how changes to the triple alpha
resonance energy affect the evolution of massive stars and their
yields of carbon, oxygen, and other elements. From a phenomenological
viewpoint, the most important variable is the energy increment
$\delres$ defined in equation (\ref{delresdef}). However, this
quantity is not a fundamental parameter of the Standard Model of
Particle Physics (see, e.g., \cite{tegmark} for one accounting of the
constants), but rather is a complicated function of the basic
parameters \cite{meissner2}. Small changes to the strength of the
nucleon-nucleon potential lead to large changes to the triple alpha
reaction rate \cite{ober1994}, primarily due to the exponential
dependence displayed in equation (\ref{trialpharate}). More
specifically, previous work has shown that variations in the resonance
energy of order $\delres\sim100$ keV correspond to variations in the
Higgs vacuum expectation value of order $\sim10\%$ \cite{jeltema}. The
requirement that stable complex nuclei exist places a weaker bound on
this expectation value \cite{agrawal}. For the same increment 
$\delres\sim100$ keV, the required changes to the nuclear potential 
and/or the electromagnetic interaction are estimated to be of order 
0.5\% and 2--4\%, respectively \cite{schlattl,ekstrom}. 

One can also assess the triple alpha constraint by considering the
energy levels of nuclei: The triple alpha reaction rate is enhanced
because the particle energies (under the relevant stellar conditions)
happen to be comparable to an excited state of the carbon nucleus.
Resonances correspond to excited states (energy levels) of the
nucleus. In general, the excited states of nuclei are spaced with
energy intervals of order $\sim1$ MeV \cite{nukereview}.  For the
specific nucleus of interest, however, five of the first excited
states of carbon have energies $E$ = 4.44, 7.65, 9.64, 12.7, and 15.1
MeV (many additional resonances are also present \cite{tunl}).  The
energy separation between adjacent resonances is $\sim3$ MeV.  The
allowed range for $\delres$ ($\sim800$ keV) corresponds to about one
fourth of the spacing interval. Given that the life-permitting range
(7.65 $\pm$ 0.3 MeV) lies in the range of carbon energy levels, the
chances for particle energies to be sufficiently near a resonance are
about 1 part in 4. Being near a resonance is neither necessary nor
sufficient for carbon production, so the odds of a successful universe
are not this favorable, but the chances of being near a resonance are
relatively high. All strong and electromagnetic nuclear reactions must
obey angular momentum and parity conservation laws, which in turn lead
to selection rules. One must also take into account the widths of the
resonances, and their proximity to other excited states, and these
considerations affect the nuclear reaction rates \cite{freer}. As a
result, a wide range of possible reactions could play a role in carbon
production in other universes, i.e., the available parameter space is
large. A full discussion of these complications is beyond the scope of
this current paper, but should should be addressed in the future.

This work has assumed that the triple alpha resonance energy varies,
but the binding energies for the relevant nuclei and the reaction
rates for other processes are unchanged. Given the allowed range for
the resonance energy from equations (\ref{resrangecar}) and
(\ref{resrangeoxy}), we can consider the consistency of this
assumption. The binding energies for carbon and helium are 92.2 MeV
and 28.3 MeV. These energies are larger than the energy increments of
interest, $|\delres|\approx300-400$ keV, by almost two orders of
magnitude. Although a fully self-consistent treatment should be
carried out in the future --- where all of the binding energies,
resonance levels, and reaction rates are re-calculated as a function
of the fundamental parameters --- the current approach provides a good
working approximation. Notice also that the nucleon binding energy and
the nuclear energy levels scale as $E\sim\alpha_s^2\mpro$
\cite{bartip}, so that the allowed range in $\delres$ corresponds to
relatively modest changes ($\sim5\%$) in the strong coupling constant
$\alpha_s$.

\section{Conclusion} 
\label{sec:conclude} 

This paper presents results from a large ensemble ($\sim2400$) of
stellar evolution simulations for massive stars in other universes
with varying values for the triple alpha resonance energy.  These
findings indicate that the sensitivity of our universe --- and others
--- to the triple alpha reaction for carbon production is more subtle
and less confining than previously reported.

\subsection{Summary of Results} 
\label{sec:summary} 

The results of our stellar evolution calculations add to our
understanding of the triple alpha fine-tuning problem, and can 
be summarized as follows: 

\medskip\noindent $\bullet$ 
The change in the triple alpha resonance energy $\delres$ is the
defining variable in the problem.  For lower values of the resonance
energy, $\delres<0$, massive stars produce more carbon than those in
our universe (not less). The allowed parameter space for viable
universes extends down to $\delres\approx-300$ keV. For larger values
of the resonance energy, $\delres>0$, the carbon yields decrease
steadily, but substantial carbon production continues up to
$\delres\approx+500$ keV. The total allowed range is $\sim800$ keV.

\medskip\noindent $\bullet$ 
Oxygen abundances decrease with decreasing $\delres$, and increase as
the resonance energy increases up to $\delres\sim150$ keV. Oxygen
abundances then decline with further increases in $\delres$. The range
of $\delres$ for which massive stars can produce significant oxygen is 
comparable to, but somewhat smaller than, that for carbon. As a result, 
the requirement that universes produce sufficient oxygen may be more 
restrictive than the constraint from carbon. 

\medskip\noindent $\bullet$ 
Although the ability of stars to produce carbon and oxygen depends
sensitively on the triple alpha resonance energy, the yields do not
show a strong dependence on stellar mass or metallicity. Similarly, 
the overarching trajectory of stellar evolution, from main-sequence 
hydrogen burning to the development of an iron core, is not greatly
affected by changes to the triple alpha process and yields of the
intermediate elements.

\medskip\noindent $\bullet$ 
The allowed range ($\sim800$ keV) for the triple alpha resonance
energy corresponds to $\sim10-20\%$ changes to the parameters of
fundamental physics. For comparison, the $^8$Be nucleus fails to be
bound by only 92 keV, and could become stable with $1-2\%$ changes to
the fundamental parameters. If $^8$Be is stable, then stars can burn
helium and make carbon through alternate reaction chains \cite{agalpha}.
The range of parameters for which the triple alpha process can operate
effectively is thus much larger than the range over which it is
necessary for carbon production.

\subsection{Discussion} 
\label{sec:discuss} 

Enforcing the requirement that other universes produce sufficient
inventories of carbon is subject to a number of complications:
Although increasing the triple alpha resonance energy leads to lower
carbon yields, as generally claimed, the allowed change $\delres$ can
be as large as $+500$ keV. On the other hand, if the resonance energy
is lower, then stars produce {\it more carbon} than in our universe,
where this finding is consistent with previous results
\cite{livio,oberhummer,schlattl}. In this regime, the increased carbon
abundances are accompanied by decreased oxygen abundances, so that a
smaller range of $\delres$ remains viable if the [C/O] ratio is
required to be of order unity \cite{tegmark,bartip,barnes2012}.
Universes produce nearly equal abundances for $\delres\sim0$, and also
for $\delres\sim+200$ keV, although the latter regime has lower
absolute abundances (see Figure \ref{fig:COfractionzero}). On another
front, the changes to nuclear physics required to compromise the
triple alpha reaction are much smaller than the changes required for
$^8$Be to become stable, which would allow stars to produce carbon
through a different set of nuclear reactions.  Yet another possible
pathway for carbon synthesis arises if the excited states of the
carbon nucleus are altered so that a different resonance is active.
Finally, one should keep in mind that the carbon abundance of Earth,
the only place in the universe where life is known to exist, is
depleted by a factor of $\sim100$ relative to cosmic and solar
abundances. These generalizations enlarge the parameter space for
viable universes.

The main result of this study is that massive stars can provide
universes with significant amounts of carbon over a wider range of
parameter space than is often considered viable. This finding adds to
a growing body of work showing that stars and stellar evolution are
relatively robust to changes in the fundamental parameters of physics
and astrophysics.  Stars exist as stable nuclear burning entities
while the fine structure constant and the gravitational constant vary
over many orders of magnitude \cite{adams,adamsnew}. Similarly, the
coefficients that set nuclear reaction rates can vary over even wider
ranges \cite{adams} and can allow stellar evolution to proceed in
universes where diprotons are stable \cite{barnes2016}. On a related
note, stars can also function in universes where deuterium is unstable
through the combined action of a triple nucleon reaction, the CNO
cycle, explosive nucleosynthesis, and power generation by
gravitational contraction \cite{agdeuterium} (cf. \cite{barnes2017}).
Stars can even operate in universes without the weak interaction
\cite{harnik}, where deuterium burning largely replaces hydrogen
burning as the principle nuclear reaction for stellar energy
generation \cite{grohsweakless}, or in universes where the weak
interaction is stronger \cite{howeweakful}. Taken together, these
results suggest that stars and stellar evolution are not the limiting
factor for universes to remain viable as the fundamental constants are
changed.

In addition to the stellar constraints of this paper, universes are
limited by additional considerations, including nuclear stability and
atomic structure. As one example, the Higgs vacuum expectation value
${\cal V}$ must fall in the range $0.39<{\cal V}/{\cal V}_0<1.64$,
where ${\cal V}_0$ is the observed value and the quark Yukawa
couplings are held constant \cite{damour}. Hydrogen becomes unstable
(through the reaction $p+e^{-}\to n+\nu$) if the expectation value
${\cal V}$ is too small, whereas nuclear binding is compromised if
${\cal V}$ is too large. For a full assessment of fine-tuning, the
relevant question is the overall size of the life-permitting region of
parameter space, which is beyond the scope of this present discussion
(e.g., see also \cite{tegmark,reessix,schellekens,carr,bartip,hogan}).

This paper extends our understanding of universes with varying values
of the triple alpha resonance energy, but a great deal of work remains
to be done. The simulations presented here show the requirements for
stars with $M_\ast=15-40M_\odot$ to produce carbon.  This mass range
can be extended to include both more massive and less massive stars,
especially intermediate mass stars with $M_\ast\approx2-10M_\odot$.
Inclusion of an extended stellar mass distribution could allow for
carbon production over an even wider range of $\delres$ than presented
herein.  In particular, the latter mass range includes progenitors
that become Asymptotic Giant Branch stars, which provide significant
contributions to the carbon inventory in our universe \cite{agbstars}.
Nucleosynthesis should also be studied in other astrophysical
environments, including supernova blast waves, white dwarf explosions,
and neutron star mergers. Carbon survives in our universe because it
is not destroyed through the reaction $^{12}$C$(\alpha,\gamma)^{16}$O,
which is non-resonant. Just as the excited states of the carbon
nucleus could be different in other universes, those of the oxygen
nucleus could also vary. Future work should consider oxygen resonances
and the ramifications for the loss of carbon.

Another important issue is to improve our understanding of the
relationship between the fundamental parameters, those that appear in
the Standard Model of particle physics, and variations in nuclear
properties. As the energy of the triple alpha resonance changes, the
reaction rates and binding energies must also vary at some level. A
self-consistent treatment remains to be carried out. On a more global
scale, the question of whether or not the universe is fine-tuned is
subject to a number of uncertainties. In spite of the extensive
literature on this topic, we still do not have a detailed model of the
multiverse, with a definitive determination of which fundamental
parameters actually vary, how such variations are coupled or
correlated, and a determination of the possible ranges of parameters
that allow for viable universes.

Although the consideration of parameter variations in other universes
is necessarily counterfactual, calculations of this type are useful in
many ways. In addition to evaluating the degree of fine-tuning
required for a universe to be habitable, another important goal is to
increase our understanding of how the universe works. For example,
this paper shows that variations in the triple alpha process lead to
different abundances of carbon and oxygen, but do not greatly change
the overall trajectory of stellar evolution. Massive stars always
start their lives with configurations capable of burning hydrogen,
then produce increasingly larger nuclei until they develop degenerate
iron cores, and end their lives with supernova explosions. Most of the
time is spent during the production of $^4$He, while the total amount
of energy generated is determined by the binding energy of $^{56}$Fe.
The inventory of intermediate elements produced along the way, while
vital to biology, is largely incidental to stellar operations. This
perspective is useful for understanding stars in our universe, whether
or not variations in the fundamental parameters are realized in other
regions of space-time.

\vskip 0.25truein
\noindent
{\bf Acknowledgments:} We would like to thank Juliette Becker, 
George Fuller, Alex Howe, Mark Paris, and Frank Timmes for useful
discussions. We also thank an anonymous referee for many comments that
improved the manuscript.  The computational resources used for this
project were provided by Advanced Research Computing--Technology
Services (ARC-TS) at the University of Michigan, where we used the
{\footnotesize{\sl FLUX}} high-performance computing cluster. This
work was supported by the University of Michigan and in part by the
John Templeton Foundation through Grant ID55112 
{\sl Astrophysical Structures in Other Universes}. EG acknowledges
additional support from the National Science Foundation, Grant
PHY-1630782, and the Heising-Simons Foundation, Grant 2017-228.

\end{document}